\documentclass[twocolumn,superscriptaddress, a4paper]{revtex4-1}
\usepackage[utf8]{inputenc}

\usepackage{graphicx}
\usepackage{ulem}

\graphicspath{{Figures/}}

\def\ket #1{\vert #1\rangle}

\usepackage{amsmath,amssymb}

\usepackage{color}
\definecolor{emeraldGreen}{rgb}{0.263, 0.608, 0.192}
\definecolor{rubyRed}{rgb}{	0.608, 0.192, 0.263}
\definecolor{triadicBlue}{rgb}{0.192, 0.263, 0.608}

\usepackage[colorlinks=true,bookmarks=false,citecolor=emeraldGreen,urlcolor=triadicBlue,linkcolor=rubyRed]{hyperref} 

\renewcommand{\emph}{\textit}

\begin{document}

\title{Simple Quantum Key Distribution with qubit-based synchronization and a self-compensating polarization encoder}

\author{Costantino Agnesi}
\thanks{These authors contributed equally to this work.}
\affiliation{Dipartimento di Ingegneria dell'Informazione, Universit\`a degli Studi di Padova, via Gradenigo 6B - 35131 Padova, Italy}
\affiliation{Istituto Nazionale di Fisica Nucleare (INFN) -- sezione di Padova, Italy}

\author{Marco Avesani}
\thanks{These authors contributed equally to this work.}
\affiliation{Dipartimento di Ingegneria dell'Informazione, Universit\`a degli Studi di Padova, via Gradenigo 6B - 35131 Padova, Italy}

 \author{Luca Calderaro}
 \thanks{These authors contributed equally to this work.}
\affiliation{Dipartimento di Ingegneria dell'Informazione, Universit\`a degli Studi di Padova, via Gradenigo 6B - 35131 Padova, Italy}
\affiliation{Istituto Nazionale di Fisica Nucleare (INFN) -- sezione di Padova, Italy}

\author{Andrea Stanco}
\affiliation{Dipartimento di Ingegneria dell'Informazione, Universit\`a degli Studi di Padova, via Gradenigo 6B - 35131 Padova, Italy}
\affiliation{Istituto Nazionale di Fisica Nucleare (INFN) -- sezione di Padova, Italy}

\author{Giulio Foletto}
\author{\\ Mujtaba Zahidy}
\author{Alessia Scriminich}
\affiliation{Dipartimento di Ingegneria dell'Informazione, Universit\`a degli Studi di Padova, via Gradenigo 6B - 35131 Padova, Italy}

\author{Francesco Vedovato}
\affiliation{Dipartimento di Ingegneria dell'Informazione, Universit\`a degli Studi di Padova, via Gradenigo 6B - 35131 Padova, Italy}
\affiliation{Istituto Nazionale di Fisica Nucleare (INFN) -- sezione di Padova, Italy}

\author{Giuseppe Vallone}
\affiliation{Dipartimento di Ingegneria dell'Informazione, Universit\`a degli Studi di Padova, via Gradenigo 6B - 35131 Padova, Italy}
\affiliation{Istituto Nazionale di Fisica Nucleare (INFN) -- sezione di Padova, Italy}
\affiliation{Dipartimento di Fisica e Astronomia, Universit\`a degli Studi di Padova, via Marzolo 8, 35131 Padova, Italy}

\author{Paolo Villoresi}
\email{paolo.villoresi@dei.unipd.it}
\affiliation{Dipartimento di Ingegneria dell'Informazione, Universit\`a degli Studi di Padova, via Gradenigo 6B - 35131 Padova, Italy}
\affiliation{Istituto Nazionale di Fisica Nucleare (INFN) -- sezione di Padova, Italy}
\email{paolo.villoresi@dei.unipd.it}

\begin{abstract}
 Quantum Key Distribution (QKD) relies on quantum communication to allow distant parties to share a secure cryptographic key. Widespread adoption of QKD in current telecommunication networks will require the development of simple, low cost and stable systems. 
  However, current QKD implementations usually include  additional hardware that perform auxiliary tasks such as temporal synchronization and polarization basis tracking.
Here we present a  polarization-based QKD system  operating at 1550~nm that performs synchronization  and polarization compensation  by exploiting only the hardware already needed for the quantum communication task.
Polarization encoding is performed by a self-compensating Sagnac loop modulator which exhibits high  temporal stability and the lowest intrinsic  quantum bit error rate reported so far.  
The QKD system was tested over a fiber-optic link,  demonstrating tolerance up to  about 40~dB of channel losses. Thanks to its reduced hardware requirements and the quality of the source, this work represents an important step towards technologically mature QKD systems.
\end{abstract}

\maketitle

\section{Introduction}

A major challenge for today's communication networks is to ensure safe exchange of sensitive data between distant parties.
However, the rapid development of quantum information protocols towards the quantum computer~\cite{Ladd2010, Flamini2019, google2019}, poses a substantial threat for current cyber-security systems. 
In fact, quantum routines such as Shor's factorization algorithm~\cite{Shor1997,Vandersypen2001,Politi2009} could potentially render today's cryptographic schemes obsolete and completely insecure.
Fortunately, Quantum Key Distribution (QKD) ~\cite{Bennett2014_BB84, Gisin_review2002, Pirandola2019rev} represents a solution to this catastrophic scenario.
By leveraging on the principles of quantum mechanics and the characteristics of photons, QKD allows two distant parties, conventionally called Alice and Bob, to distill a perfectly-secret key and bound the shared information with any adversarial eavesdropper~\cite{Scarani2008}.
Furthermore, QKD is an interesting solution for applications requiring long term privacy since algorithmic and technological advances for both classical and quantum computation do not threaten the security of keys generated with QKD. 

\begin{figure*}[t]
\centering
\includegraphics[width=\linewidth]{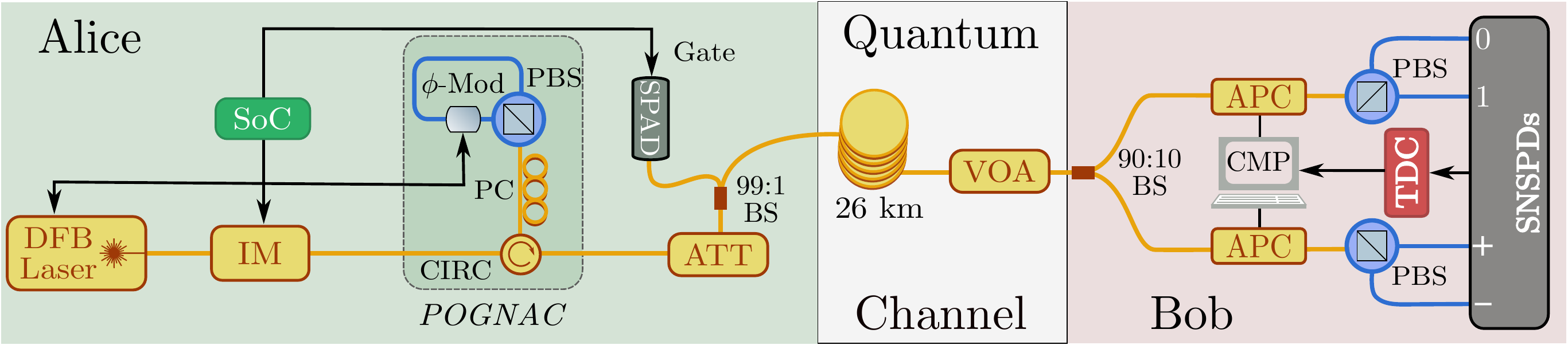}
\caption{\textbf{Experimental Setup.} For a detailed description  see Sec.~\ref{section:Setup}. Single Mode fibers are indicated in yellow while Polarization Maintaining fibers in blue.}
\label{fig:setup}
\end{figure*}

Since its first proposal by Bennet and Brassard in 1984~\cite{Bennett2014_BB84}, QKD has received much attention and several experiments have shown its feasibility by exploiting different photonic degrees of freedom and platforms in free-space~\cite{Erven:08, Vallone2014, Liao2017_Day, QCOSONE}, optical fibers~\cite{Dixon:08, korzh2015, Islame1701491, Sibson:17, Boaron2018, Bunandar2018}, or even satellite links~\cite{Vallone2015prl, vallone2016prl, Liao2017_Sat, Bedington2017, Agnesi2018}. Recent developments have focused mainly in rendering QKD implementations simpler and more robust, aiming for compatibility with standard communication networks and widespread usage. This has led, for example, to the introduction of self-compensated modulators for different photonic degrees of freedom such as time-bin~\cite{Wang2018}, mean photon number~\cite{Roberts2018}, and polarization~\cite{Agnesi2019,Li:19}, all based on Sagnac interferometric configurations. Also, simpler QKD protocols have been introduced such as a three-state~\cite{Fung2006, Tamaki2014} and one-decoy state version of the BB84 protocol, which simplifies the requirements of the quantum state encoder~\cite{Grunenfelder2018} and can provide higher rates in the finite-key scenario~\cite{Rusca2018}.

A critical aspect of QKD systems is the distribution of a temporal reference between the transmitter (Alice) and the receiver (Bob). This is crucial for at least two reasons. First, it allows to discriminate between the quantum signal and the  noise introduced by either the quantum channel or detector defects. Secondly, it allows to correlate the qubit sequence transmitted by Alice with the detection events recorded by Bob. This correlation enables the distillation of the quantum-secure cryptographic key. The transmission of the temporal reference is usually achieved by sending a decimated version of Alice's clock  using an additional laser communication system, that, in turn, requires the use of a secondary fiber channel~\cite{Liu2010, korzh2015}, or time or wavelength multiplexing schemes to separate the quantum information from the classical light pulses~\cite{Liao2017_Sat,Treiber_2009}. Also, Global Navigation Satellite Systems (GNSS) can be used to synchronize Alice and Bob since these systems can give precise temporal references~\cite{Vallone2015,Bourgoin2015, QCOSONE}. All these approaches, however, require additional hardware with respect to what is already needed  for the quantum communication task.

Polarization-encoded QKD in fiber-optic links has been studied to great extent~\cite{Peng2007,Treiber_2009,Liu2010,Grunenfelder2018,Li2018}.
Unfortunately, this type of link has an important drawback given by the natural birefringence of optical fibers, which causes the polarization state of transmitted photons to change continuously and in an unpredictable fashion~\cite{Ding2017}. Several approaches have been conceived to counteract these random polarization drifts, most of them requiring auxiliary laser pulses and  time or wavelength multiplexing schemes~\cite{Peng2007,Treiber_2009,Xavier2008,Li2018}, which,  similarly to the synchronization task discussed above, require additional hardware. A different approach was introduced by Ding \textit{et al.} that used the revealed portion of the sifted key~\cite{Ding2017_tracking}, produced during the error correction and privacy amplification procedures, to detect and compensate the polarization drifts of the fiber link. Unfortunately, this method requires post-processing of an entire block of raw key, imposing a limit to the polarization tracking speed.

 Here we present a simple QKD system, in which quantum communication, temporal synchronization and  polarization compensation are all realized in the same optical setup. 
The temporal synchronization is performed using a novel method, whose technical details are presented elsewhere~\cite{Calderaro2019}. 
This  method does not require any auxiliary time reference and works by sending a public qubit sequence at pre-established times. Hence, it is named {\it Qubit4Sync}, because it uses only qubits for synchronization.
Predetermined qubit sequences are also exploited to monitor and compensate the polarization drift introduced by the quantum channel, constituted by a 26 km-long fiber spool.
With respect to Ref.~\cite{Ding2017_tracking}, our solution does not require post-processing of an entire block of raw key, allowing for an increased compensation speed.
Furthermore, the QKD source here presented exhibits several hours of stability and an intrinsic quantum bit error rate (QBER) of the order of 0.05\%, which is, to the best of our knowledge, the lowest reported so far. This source exploits the scheme for polarization-encoding based on a Sagnac loop (hence the name {\it POGNAC}) we introduced in Ref.~\cite{Agnesi2019}.
The relaxed  hardware requirements, the high stability and the record-low QBER of the presented implementation represent an important technological step towards mature and efficient QKD systems.

\section{Setup}
\label{section:Setup}

 Our experimental setup, that implements the simplified three-state and one-decoy protocol proposed in~\cite{Grunenfelder2018}, is sketched in Fig.~\ref{fig:setup}.
A gain-switched distributed feedback (DFB) laser source outputs a 50~MHz stream of phase-randomized pulses with 270~ps of full-width-at-half-maximum temporal duration at 1550 nm wavelength. The light pulses first pass through a Lithium Niobate intensity modulator (IM) used to set the intensity levels required by the decoy-state method.
The pulses then enter the {\it POGNAC} polarization modulator realized using only standard commercial off-the-shelf (COTS) fiber components, namely a circulator (CIRC), a polarization controller (PC), a polarizing beam splitter (PBS) and a phase modulator ($\phi$-Mod).  Compared to other polarization modulators, the {\it POGNAC} requires a lower $V_\pi$ voltage, can be developed using $\phi$-Mods that carry only a single polarization mode, exhibits no polarization mode dispersion, and has a self-compensating design that guarantees robustness against temperature and electrical drifts (see Ref.~\cite{Agnesi2019} for a full description and additional details).

The photons emerge from the {\it POGNAC} with a polarization state given by 
\begin{equation}
 \ket{\psi^{\phi_e,\phi_\ell}_\mathrm{out}}  = \frac{1}{\sqrt{2}} \left( \ket{H} +  e^{i(\phi_e-\phi_\ell)} \ket{V} \right) ,
\end{equation}
where the phases $\phi_e$ and $\phi_\ell$ can be set by carefully timing the applied voltage on a Lithium Niobate $\phi$-Mod. This was achieved with the Zynq-7000 ARM/FPGA System-on-a-Chip (SoC, manufactured by Xilinx), which in our implementation controls the operation of the QKD source.

If no voltages are applied by the SoC, the polarization state remains unchanged, {\it i.e.}
$\ket{+} = \left( \ket{H} + \ket{V} \right)/\sqrt{2}$.
Instead, if $\phi_e$ is set to $\frac{\pi}{2}$ while $\phi_\ell$ remains zero, the output state becomes 
$\ket{L} =  \left( \ket{H} + i \ket{V} \right)/\sqrt{2}$.   
Alternatively, if $\phi_e$ remains zero  while $\phi_\ell$ is set to $\frac{\pi}{2}$, the output state becomes
$\ket{R} =  \left( \ket{H} - i \ket{V} \right)/\sqrt{2}$.
In this way we generate the three states required by the simplified three polarization state version of BB84~\cite{Grunenfelder2018}, with the key-generation basis $\mathcal{Z} = \{\ket{0}, \ket{1}\}$ where $\ket{0}:=\ket{L}$, $\ket{1}:= \ket{R}$, and the control state $\ket{+}$ of the $\mathcal{X}=  \{\ket{+}, \ket{-}=\left( \ket{H} - \ket{V} \right)/\sqrt{2}\}$ basis.

The optical pulses then encounter an  optical attenuator (ATT) which weakens the light to the single photon level. 
A 99:1 beam splitter (BS) is used to estimate the intensity level of the pulses: the  1\%  output port is directed to a gated InGaAs/InP  Single Photon Avalanche Diode (SPAD, manufactured by Micro Photon Devices Srl~\cite{Tosi2012}), while the other output port is directed to the quantum channel (QC).
In our implementation the QC is formed by a 26~km spool of G.655 dispersion-shifted fiber with $0.35$~dB/km of loss followed by a variable optical attenuator (VOA). This VOA allows us to introduce further channel loss in order to test our system's resilience.

Alice sends key-generation states with probability $p_A^\mathcal{Z} = 0.9$ ($p_A^\mathcal{X} = 0.1$), while the two intensity levels are $\mu_1 \approx 0.80$ and $\mu_2 \approx 0.28$, which are sent with probabilities $p_{\mu_1} = 0.7$  and $p_{\mu_2} = 0.3$ respectively.  This intensity modulation is driven by the SoC, and presents no drifts during the QKD runs, as attested by the data presented in the Supplementary Material (SM). The used parameters are close to optimal according to our simulations and Ref.~\cite{Rusca2018}.
The random bits used in the QKD runs are obtained from the Source-Device-Independent quantum random generator based on optical heterodyne measurements described in Ref.~\cite{Avesani2018}.

The fiber receiving setup consists of a 90:10 fiber BS setting the detection probabilities of the two measurement bases to $p^\mathcal{Z}_B = 0.9$ and $p^\mathcal{X}_B = 0.1$. Each output arm of the BS is connected to an automatic polarization controller (APC) and a PBS. The four outputs are then sent to four superconductive nanowire single-photon detectors (SNSPDs, manufactured by ID Quantique SA) cooled to 0.8~K. The detection efficiencies are around 85\% for the detectors in the $\mathcal{Z}$ basis, whereas it is 90\% and 30\% for the $\ket{+}$ and $\ket{-}$ detectors, respectively. As discussed in Refs.~\cite{Bochkov2019,QCOSONE}, some events are randomly discarded in post-processing to balance the different efficiencies. 
All the detectors are affected by about 200~Hz of free-running intrinsic dark count rate.
The SNSPD detections are recorded by the quTAG time-to-digital converter (TDC, manufactured by qutools GmbH) with 1~ps of temporal resolution and jitter of 10~ps.
A computer (CMP) then reads the TDC data and uses it for temporal synchronization, polarization compensation and QKD. The low dark count rate and negligible afterpulsing represent the main reasons that made us choose SNSPDs over, for example, InGaAs SPADs. Indeed, a low dark count rate allows the QBER to stay low even for strong levels of channel attenuation ({\it i.e.} long fiber links).

\subsection{Synchronization}
\label{subsection:sync}
In this work, we use the {\it Qubit4Sync} algorithm to synchronize Alice and Bob's clocks using the same qubits exchanged during the QKD protocol. This means that the setup does not need any synchronization subsystem, which is usually implemented with a pulsed laser or GNSS clock to share an external time reference. 
The synchronization method is described in detail in Ref.~\cite{Calderaro2019}. 

Here we report the main features of the algorithm. The synchronization is done in post-processing, adjusting the times in which Bob expects to receive the qubits from Alice. For this, Bob needs to determine at which frequency (in his time reference) the qubits are arriving at the detectors and the absolute time in which the first qubit should arrive. Our approach is to compute the frequency from the time-of-arrival measurements. To recover the absolute time, we send an initial public string encoded in the first $L$ states. By correlating this string with the one received by Bob, it is possible to distinguish which state received by Bob is the first one sent by Alice, hence the absolute time of the first qubit. This is the typical technique used for instance by the GPS receiver to synchronize with the satellite signal~\cite{Hassanieh2012}. 

The novelty of {\it Qubit4Sync} is the implementation of a fast correlation algorithm requiring lower computational cost than the algorithms based on a sparse fast Fourier transform, as we show in Ref.~\cite{Calderaro2019}. This allows us to calculate, in real-time, the position of the maximum correlation peak of long synchronization strings, which is required to cope with the high losses of a quantum channel. To the best of our knowledge, no similar algorithms have been previously proposed or used for QKD.

\subsection{Polarization compensation scheme}
\label{subsection:PolComp}
Mechanical and temperature fluctuations lead to variations in the natural birefringence of fiber optics, transforming the polarization state of the photons that travel through the fiber.
This transformation is troublesome for QKD since it causes Alice and Bob to effectively have different polarization reference frames.
As a consequence of this mismatch the QBER increases, lowering the Secure Key Rate (SKR) up to the point where no quantum secure key can be established. To prevent this a polarization compensation system must be utilized. 

Here we propose a polarization compensation scheme that exploits a shared public string, not necessarily related to the synchronization string. Every second, the shared string of $10^6$ states is transmitted by Alice encoded using weak coherent pulses in the $\mathcal{Z}$ basis with $\mu_1$ intensity.  Bob detects the sequence and after performing the temporal synchronization routine he estimates the QBER of his recorded sequence.
Bob still has to estimate the $\mathcal{X}$ basis QBER. For this purpose, at the end of each interval Alice reveals the basis used to encode the QKD qubits that follow the public string. This process is actually the standard  basis reconciliation procedure of QKD. Since in this protocol only one state is transmitted in the $\mathcal{X}$ basis, Bob can immediately estimate the QBER~\cite{Grunenfelder2018}.

The estimated QBER values are then fed into an optimization algorithm,  based on coordinate descent~\cite{Wright2015} and running in Bob's CMP, which controls the APCs of Bob's setup.
The APCs have 4 different piezoelectric 1-D actuators, alternately at $0^\circ$ and $45^\circ$ to the horizontal plane, that stress and strain the optical fibers, changing the polarization of the light that traverses them~\cite{Walker1987}. Our optimization algorithm loops through the 4 actuators sequentially. At each round, the position of an actuator is changed with a step size  proportional to the measured QBER. If such change causes a reduction in the measured QBER, our algorithm keeps changing the position of the same actuator in the same direction, always with a step size proportional to the measured QBER. Instead, if an increased QBER is measured the algorithm reverses the direction of motion for the actuator. Only one reversal is permitted per round, after which the next actuator is selected and a new round begins. 

Compared to  Ref.~\cite{Ding2017_tracking}, our approach has the advantage that only the basis reconciliation step is required to obtain sufficient information to run the polarization compensation algorithm. This renders our approach less communication-intensive, and we were able to achieve a 1 second feedback cycle, which is 12 times faster than the one reported in Ref.~\cite{Ding2017_tracking}. Also, the length of the shared string and its transmission frequency can be changed to best match the requirements of the fiber optical link. Furthermore, the public string can be transmitted in an interleaved fashion together with the QKD qubits at predetermined times.

\section{Results}
\label{section:Results}

\subsection{ POGNAC low intrinsic QBER and high stability}
 The intrinsic   (or ``optical'') quantum bit error rate QBER$_{\rm opt}$ of the source gives a quantitative and qualitative measure of its  suitability for use in QKD~\cite{Gisin_review2002}.  Its characterization is relevant to predict the SKR under different conditions, such as different channels or detector technologies. It is also meaningful to measure its stability to find how long the source can function without realignment. 

\begin{figure}[t]
\centering
\includegraphics[width=\linewidth]{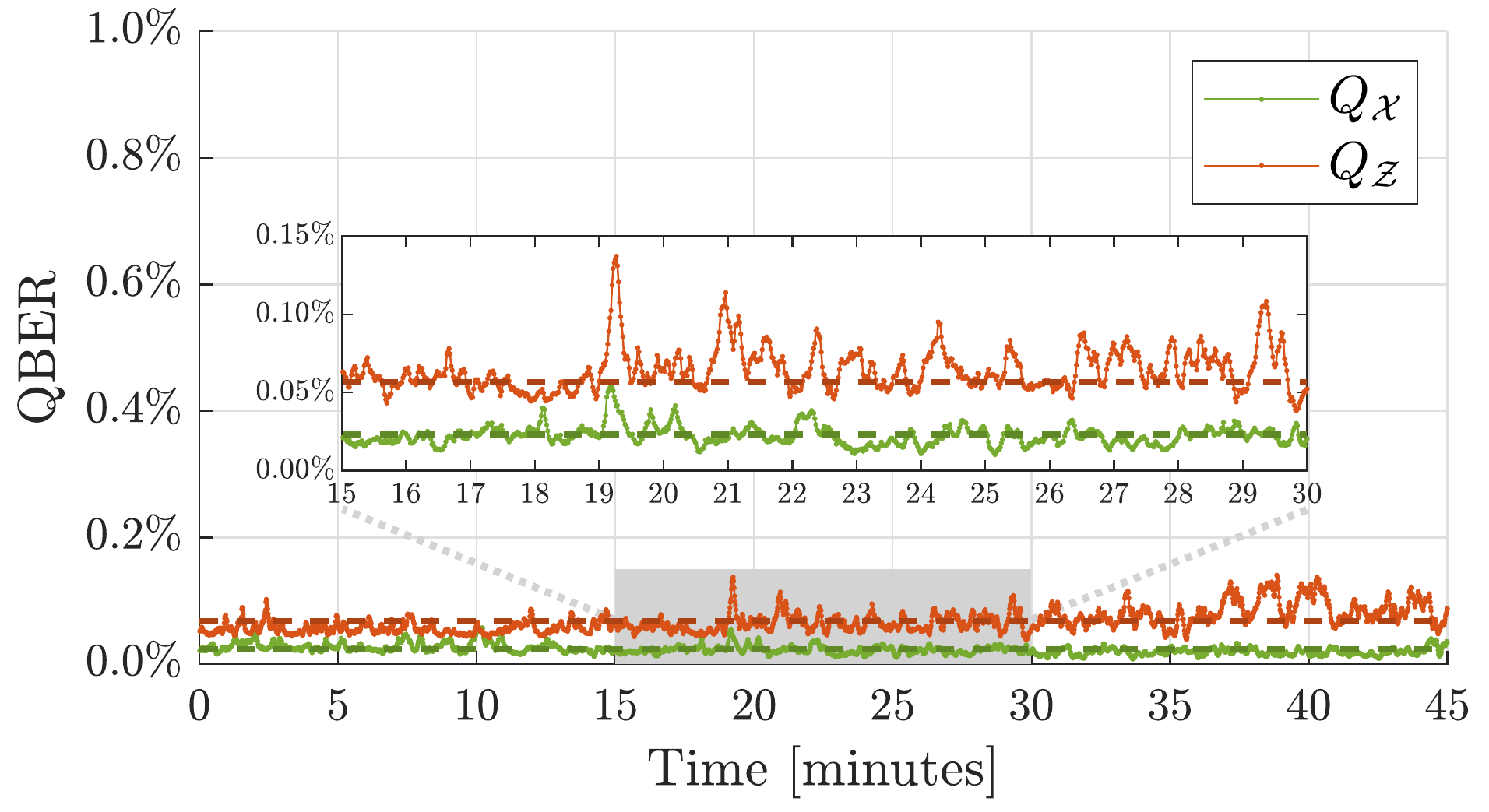}
\caption{ \textbf{Intrinsic QBER and temporal stability of the  POGNAC polarization encoder. } The average QBER measured for the key-generation basis was $Q_{\mathcal{Z}} = 0.07 \pm 0.02 \%$ (dashed red line) while an average $Q_{\mathcal{X}} = 0.02 \pm 0.01 \%$ (dashed green line) was measured for the control basis. A close-up between minutes 15 to 30 can be seen in the inset plot. }
\label{fig:IntrinsicStability}
\end{figure}

 For these reasons, we report in Fig.~\ref{fig:IntrinsicStability}, the  stability of  the intrisic QBER of our QKD polarization source.
This measurement was performed by sending a pseudo-random qubit
 sequence of $\{ \ket{0},  \ket{1},  \ket{+} \} $ states (used only for this test) and measuring the QBER of the sifted string recovered by Bob.
To remove all fluctuations not attributable to the source, the  fiber spool of the QC was bypassed  while the VOA was set to $\approx$~11 dB of attenuation.
Furthermore, the 90:10 BS was replaced with a 50:50 BS in order to have comparable statistics for both measurement bases.
Every second the QBER was estimated for both the $\mathcal{Z}$ key-generation basis and the $\mathcal{X}$ control basis.
In 45 minutes an average QBER of $Q_{\mathcal{Z}} = 0.07 \pm 0.02 \%$ was measured for the $\mathcal{Z}$ basis while the average QBER for the $\mathcal{X}$ was $Q_{\mathcal{X}} =  0.02 \pm 0.01 \%$, giving a mean QBER$_{\rm opt}$ in the two relevant bases for QKD of 0.05\%. This corresponds, to an extinction ratio of  33~dB  for the used states. We note that the reported data include the contribution from dark counts, and therefore slightly overestimate the value of the QBER$_{\rm opt}$.
To verify that the $\mathcal{X}$ and $\mathcal{Z}$ bases are mutually unbiased, we also measured the QBER of the $\mathcal{Z}$ states when observed in the $\mathcal{X}$ basis and viceversa, obtaining $48.8 \pm 0.4 \%$ in both cases.

These measurements corroborate the results of Ref.~\cite{Agnesi2019} and demonstrate  the low intrinsic  QBER, and high stability of the {\it POGNAC} polarization modulator. 
 It is worth noticing that an extinction ratio above 30~dB is typically not achievable by using COTS polarization modulators, which  also suffer from temporal drifts due to temperature and electronics fluctuations. These drifts can be suppressed by exploiting self-compensating schemes, as the {\it POGNAC} or the one in Ref.~\cite{Martinez2009}. However, Ref.~\cite{Martinez2009} reported a limited extinction ratio of less than 20~dB due to implementation imperfections.

 Table \ref{tab:comparison_qberopt} reports a comparison of the intrinsic QBER with the existing literature, in particular Refs.~\cite{Honjo:04,Gordon2004, Bunandar2018, Li:19, Boaron_APL2018, zbinden_lowqber,roberts2017, QCOSONE}. The QBER$_{\rm opt}$ we registered here is the lowest ever reported, even considering encodings other than polarization (as used in Refs.~\cite{Gordon2004, Bunandar2018, Li:19, QCOSONE}), such as time-bin (in Refs.~\cite{Boaron_APL2018, zbinden_lowqber,roberts2017}) and differential phase shift (in Ref.~\cite{Honjo:04}), as well as different platforms to realize the source, as fibers-based schemes (in Refs.~\cite{Honjo:04, Gordon2004, Li:19, zbinden_lowqber, roberts2017}) or integrated photonics chips (in Refs.~\cite{Bunandar2018, QCOSONE}).

\begin{table}[h]
    \centering
\begin{tabular}{c|c|c|c}
         Ref. & QBER$_{\rm opt}$& Encoding & Notes  \\ \hline\hline
         \cite{Honjo:04} & 0.46\%  & DPS & Estimated via ER \\      \cite{Gordon2004} & 0.4\% & Pol & Measured \\
          \cite{Bunandar2018} & 0.3\% & Pol & Estimated via ER \\
          \cite{Li:19} & 0.27\% & Pol & Measured \\
          \cite{Boaron_APL2018} & 0.25\% & TB & Estimated via $\mathcal{V}$\\
         \cite{zbinden_lowqber} & 0.15\% & TB & Estimanted via $\mathcal{V}$ \\
          \cite{roberts2017} & 0.1\% & TB & Estimated via $\mathcal{V}$ \\
          \cite{QCOSONE} & 0.1\% & Pol & Estimated via ER \\
          This work & 0.05\% & Pol & Measured\\
    \end{tabular}
    
    \caption{{\bf Comparison between intrinsic QBERs reported in literature.}  If the intrinsic extinction ratio 
    (ER) 
    of the source is provided, QBER$_{\rm opt}$ is estimated 
    via
    ${\rm QBER}_{\rm opt} = {\rm ER}/(1+{\rm ER})$~\cite{Honjo:04}. If the intrinsic fringe-visibility $\mathcal{V}$ is measured, then ${\rm QBER}_{\rm opt} = (1-\mathcal{V})/2$~\cite{Gisin_review2002}.  We include QKD sources with different encodings: differential phase shift (DPS), polarization (Pol), and time-bin (TB). }
    \label{tab:comparison_qberopt}
\end{table}

\subsection{Polarization drift compensation with 26~km of optical fiber}

\begin{figure}[b]
\centering
\includegraphics[width=\linewidth]{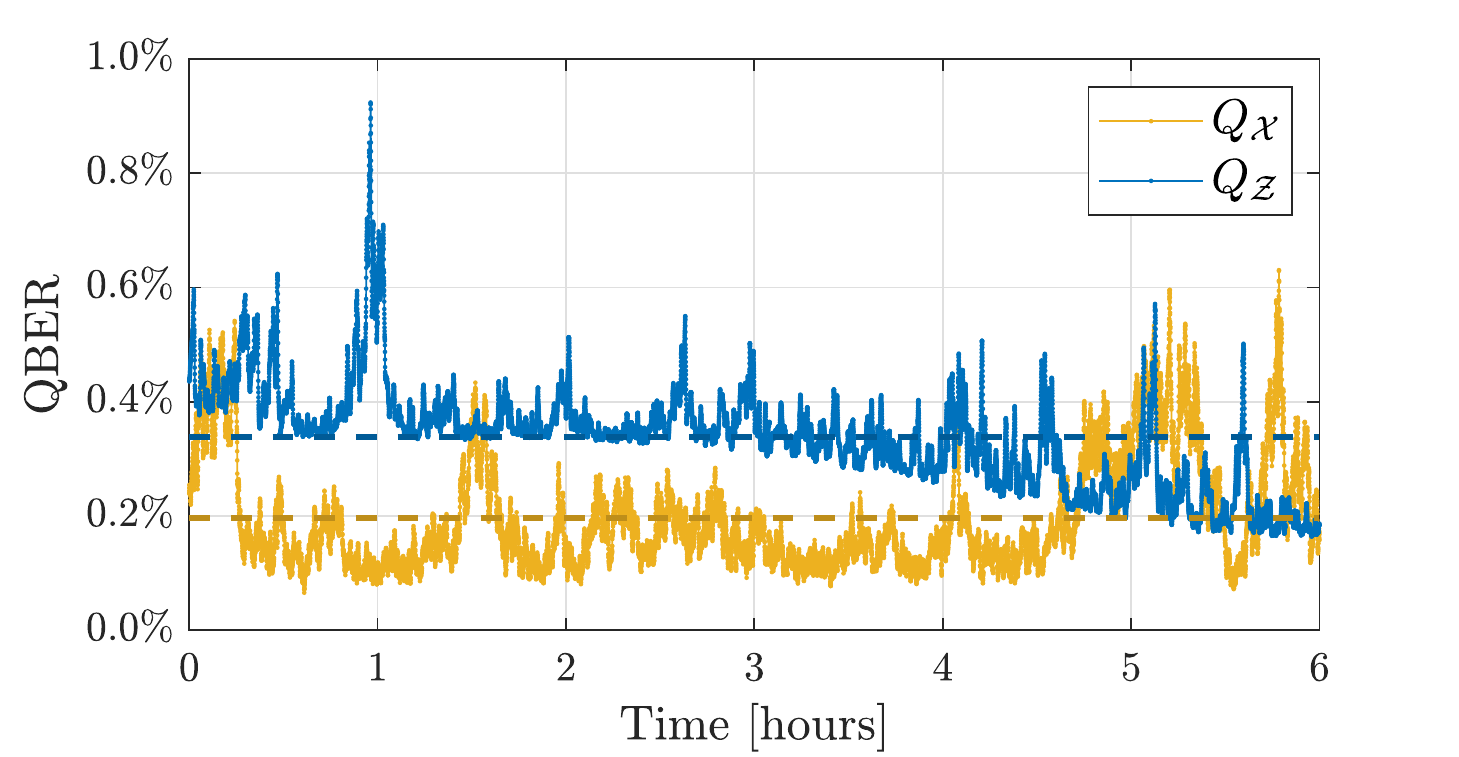}
\caption{\textbf{QBER measurement for a 6 hour long acquisition along a 26~km optical fiber channel.}  The average QBER measured for the key-generation basis was $Q_{\mathcal{Z}} = 0.3 \pm 0.1 \%$ (dashed blue line) while an average $Q_{\mathcal{X}} = 0.2 \pm 0.1 \%$ (dashed yellow line) was measured for the control basis. }
\label{fig:LongRun}
\end{figure}

To test our polarization drift compensation algorithm we performed a 6 hour long run with the QC including both the 26~km optical fiber spool  and the VOA for $\approx$~19~dB of total losses.
On average, the detected bits of the shared polarization compensation string in the $\mathcal{Z}$ basis were $\approx 8 \times 10^3$ while the sifted bits from the control basis were $\approx 3 \times 10^3$.
This allowed to correct the polarization drift with an average QBER measured for the key-generation basis of $Q_{\mathcal{Z}} = 0.3 \pm 0.1 \%$ while an average $Q_{\mathcal{X}} = 0.2 \pm 0.1 \%$ for the control basis, for six hours of continuous operation  (see Fig.~\ref{fig:LongRun}).
 These values are about an order of magnitude lower than those observed in Ref.~\cite{Ding2017_tracking}.

After the experimental run, we noted a lower detection efficiency of 45\% for the detectors of the $\mathcal{Z}$ basis. This was due to a non-optimal polarization rotation of the photons entering the SNSPD detectors, which are polarization sensitive. This reduced detection efficiency did not hamper the polarization drift compensation algorithm demonstrating its robustness even in non-optimal conditions. 

\subsection{QKD secure key rate for different channel losses}

\begin{figure}[tbp]
\centering
\includegraphics[width=\linewidth]{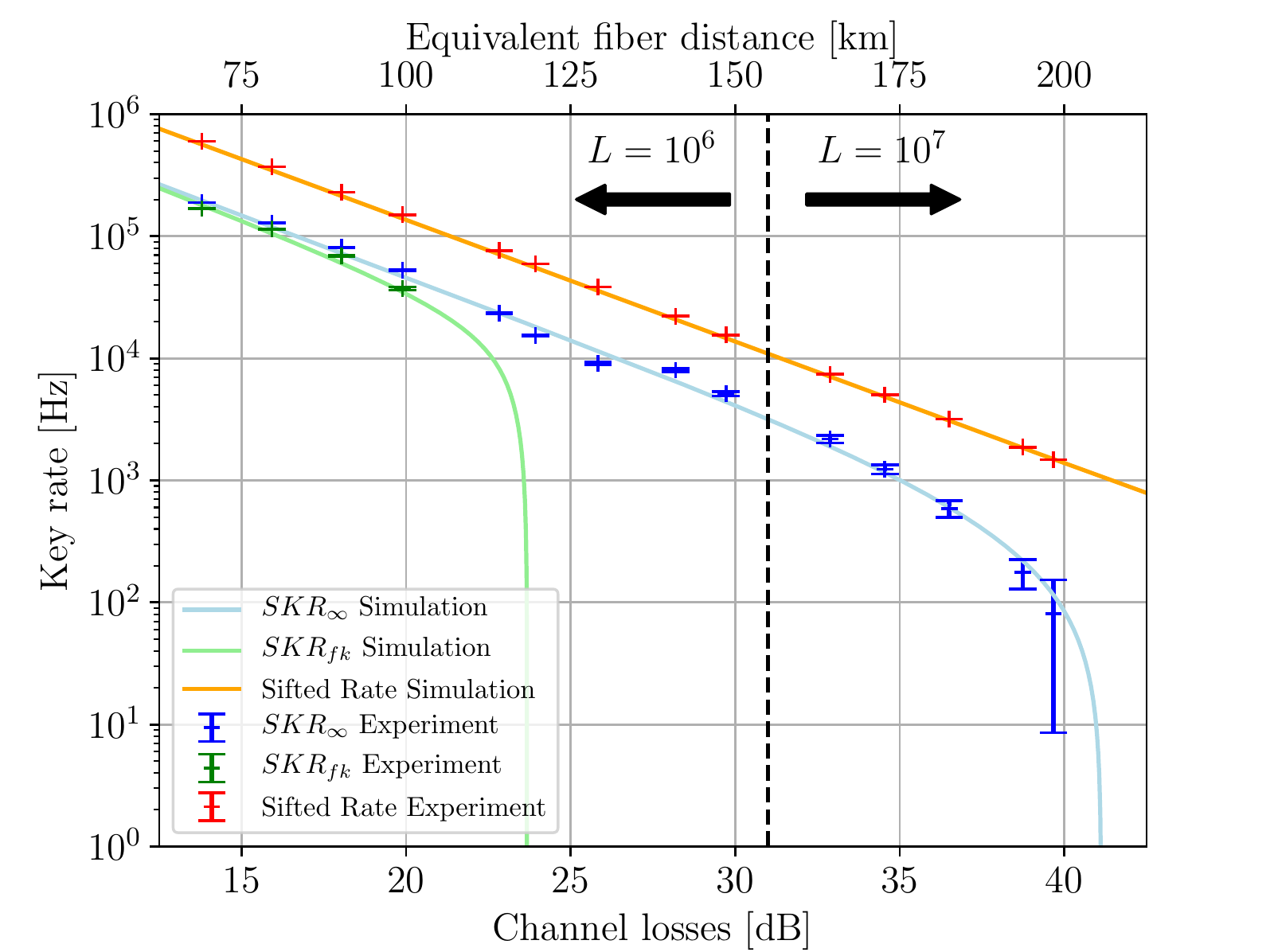}
\caption{\textbf{Sifted and secure key rate as a function of channel losses.}  For the four runs with lower losses we also include finite-key analysis ($SKR_{fk}$), for 90 seconds of acquisition each. The equivalent fiber distance (upper $x$-axis) is based on SMF28 losses (0.2 dB/km). The crosses represent the experimental runs, while the lines show the results of our simulation based on the physical parameters of our experiment. Error bars are standard deviations, obtained by simulating 1000 repetitions of the experiment. }
\label{fig:SKRloss}
\end{figure}

To test the performances of our system with qubit-based synchronization and self-compensating polarization encoder, as well as its resistance to channel losses, several  QKD runs were executed each with increased losses, as reported in Fig.~\ref{fig:SKRloss}. The losses were added increasing the attenuation of the VOA after the 26~km of fiber. A random qubit sequence of $\{ \ket{0},  \ket{1},  \ket{+} \}$ states was transmitted at a repetition rate of 50~MHz, where the first $L$ qubits of the sequence formed the publicly known synchronization string. 

For each run the SKR was calculated in the asymptotic limit according to
\begin{equation}
     SKR_\infty = \left[ s_{\mathcal{Z},0} + s_{\mathcal{Z},1}(1 - h(\phi_{\mathcal{Z}})) - f\cdot n_{\mathcal{Z}}\cdot h(Q_{\mathcal{Z}})\right]/t \ ,
\end{equation}
where $t$ is the duration of each acquisition, $h(\cdot)$ is the binary entropy, $f  = 1.06$ is the Shannon inefficiency of typical error correction algorithms,  $n_{\mathcal{Z}}$ is the length of the sifted key in the $\mathcal{Z}$ basis, $s_{\mathcal{Z},0}$ and $s_{\mathcal{Z},1}$ are the lower bounds on the number of vacuum and single-photon detections in the $\mathcal{Z}$ basis,  and $\phi_{\mathcal{Z}}$ is the upper bound on the phase error in the $\mathcal{Z}$ basis calculated from  $Q_{\mathcal{X}}$ as in Ref.~\cite{Rusca2018}, but without finite-key corrections.

 For the four runs with lower losses, we also performed the finite-key analysis using the bits produced in $t= 90$~s of acquisition by using~\cite{Rusca2018}
\begin{equation}
    SKR_{fk} = SKR_\infty - \left[6\log_2(19/\epsilon_{\rm sec}) + \log_2(2/\epsilon_{\rm conf})\right]/t \ , 
\end{equation} 
 where $s_{\mathcal{Z},0}$, $s_{\mathcal{Z},1}$ and $\phi_{\mathcal{Z}}$ in $SKR_{\infty}$ now include finite-key corrections, and with secrecy and confirmation of correctness parameters $\epsilon_{\rm sec} = 10^{-10}$ and $\epsilon_{\rm conf} = 10^{-15}$, respectively.  In the SM we also included simulations of the finite-key performance of the system with different key sizes and duration. As shown there, the system is able to produce a positive $SKR_{fk}$ for up to $\approx 38$~dB of channel losses with an acquisition time $t = 6$ hours, compatible with the measured stability of Fig.~\ref{fig:LongRun}.

As discussed in Ref.~\cite{Calderaro2019}, if the background and dark counts are not considered, the synchronization can be established  with $L=10^6$ for up to 40~dB of total losses, {\it i.e.} considering channel and receiver losses as well as  detector inefficiency.  A longer string, with $L=10^7$, could be used to synchronize up to 50~dB of losses. In our experiment, the presence of dark counts lowers the bounds by about 6~dB. Indeed, using a synchronization string of length $L=10^6$, we performed several QKD runs with losses up to 34~dB. With $L=10^7$, we successfully ran QKD  up to the channel loss at which the key rate drops to zero. In the QKD run with highest losses, we achieved a secure key rate of 80 bits per second at  40~dB channel losses, corresponding to about  200~km of SMF28 fiber (0.2~dB/km) or  235 of ultralow-loss fiber (0.17~dB/km). It is important to note that our QKD implementation withstands up to  41 dB of channel loss, as reported in the $SKR_\infty$ simulation of Fig.~\ref{fig:SKRloss}. Our results prove that the {\it Qubit4Sync} method properly works even at the highest losses tolerated by our QKD implementation.

\section{Conclusions}

Here, we have presented a simple polarization encoded QKD implementation with qubit-based synchronization and a self-compensating polarization modulator. 
Its simple  and hardware-efficient design reduces the complexity for both the QKD transmitter and receiver. In fact, the same optical setup is used for three different tasks, {\it i.e.} synchronization, polarization compensation and  quantum communication, without requiring any changes to the working parameters of the setup or any additional hardware. 
 The QKD transmitter shows high stability and an intrinsic QBER below 0.1\%. This, in addition to the effective polarization compensation technique and the use of high-performance SNSPD detectors, allows us to obtain high SKRs and resilience up to about 40~dB of channel losses, even with a repetition-rate of 50~MHz. Indeed, although the repetition rate of our source is an order of magnitude smaller than those of recent polarization-encoded fiber-based QKD experiments~\cite{Sibson:17,Bunandar2018}, we achieved a SKR that is comparable with that of Ref.~\cite{Bunandar2018} and an order of magnitude higher that reported in Ref.~\cite{Sibson:17} for distances greater than 50 km.
If the SNSPDs were replaced with InGaAs SPADs, with a free-running dark count rate of 500~Hz, 15\% quantum efficiency and 20~$\mu$s hold-off time~\cite{Tosi2012}, we expect that the system would be able to produce a positive $SKR_\infty$ for up to 35~dB (31~dB) of channel losses using a temporal gating window of 0.3~ns (1~ns).

 Currently, the {\it POGNAC} requires to be manually aligned once everyday.  However, to make our system more autonomous, its PC could be replaced with an APC controlled by a power monitor inside the fiber Sagnac loop. This would render our implementation compatible with different  operative scenarios, ranging from urban QKD fiber links~\cite{Bunandar2018} to free-space satellite QKD links via CubeSats~\cite{Oi2017}, or even to implement other quantum communication schemes such as Quantum Digital Signatures~\cite{An:19} or Remote Blind Qubit Preparation~\cite{Jiang2019}.
Lastly,  our implementation is particularly promising for free-space QKD~\cite{Liao2017_Day,Liao2017_Sat,QCOSONE} since polarization is not significantly affected by atmospheric propagation~\cite{Bonato2006} and long term stability is required, especially for links with satellites in Medium Earth Orbit~\cite{Dequal2006} or part of a GNSS constellation~\cite{Calderaro2018}.

\section*{Funding Information}

Ministero dell'Istruzione, dell'Universit\`a e della Ricerca (MIUR),  ``Fondo per il finanziamento dei dipartimenti universitari di eccellenza'' 
(Legge 232/2016) (``Internet of things: sviluppi metodologici, tecnologici e applicativi'');
Agenzia Spaziale Italiana (ASI), ``Q-SecGroundSpace'' (E16J16001490001);
Istituto Nazionale di Fisica Nucleare (INFN), ``MoonLIGHT-2''.
M.~Z. acknowledges funding from the European Union's Horizon 2020 research and innovation programme under the Marie Sk\l{}odowska-Curie grant agreement No 675662.

\section*{Acknowledgments}
The authors thank L.~Palmieri and M.~Calabrese for technical support on the 26km fiber spool.

\bibliographystyle{apsrev4-1}
\bibliography{References}

\end{document}